\documentclass[aps,prd,superscriptaddress,12pt,showpacs,notitlepage]{revtex4}
\usepackage{amsmath,amssymb,mathrsfs}
\usepackage{graphicx}

\newcommand{\be}{\begin{equation}}
\newcommand{\ee}{\end{equation}}
\newcommand{\bea}{\begin{eqnarray}}
\newcommand{\eea}{\end{eqnarray}}

\newcommand{\bb}{\bibitem}

\begin{document}
\title{Topological energy bounds in generalized Skyrme models}
\author{C. Adam}
\affiliation{Departamento de F\'isica de Part\'iculas, Universidad de Santiago de Compostela and Instituto Galego de F\'isica de Altas Enerxias (IGFAE) E-15782 Santiago de Compostela, Spain}
\author{A. Wereszczynski}
\affiliation{Institute of Physics,  Jagiellonian University,
Reymonta 4, Krak\'{o}w, Poland}

\pacs{11.27.+d, 12.39.Dc}

\begin{abstract}
The Skyrme model has a natural generalization amenable to a standard hamiltonian treatment, consisting of the standard sigma model and the Skyrme terms, a potential, and a certain term sextic in first derivatives. Here we demonstrate that, in this theory, each pair of terms in the static energy functional which may support topological solitons according to the Derrick criterion (i.e., each pair of terms with opposite Derrick scaling) separately posesses a topological energy bound. As a consequence, there exists a four-parameter family of topological bounds for the full generalized Skyrme model. The optimal bounds, i.e., the optimal values of the parameters, depend both on the form of the potential and on the relative strength of the different terms. It also follows that various submodels of the generalized Skyrme model have one-parameter families of topological energy bounds. We also consider the case of topological bounds for the generalized Skyrme model on a compact base space as well as generalizations to higher dimensions.
\end{abstract}

\maketitle 

\section{Introduction}
The Skyrme model \cite{skyrme1} - \cite{skyrme3} is a nonlinear field theory of an SU(2) valued scalar field $U$. Physically, the field variables are interpreted as pions, and the theory is regarded as an approximate low-energy effective field theory for QCD. One salient feature of the model is the presence of a topological lower bound for the energy, and the intimately related existence of topological soliton solutions ("skyrmions") \cite{skyrmions1} - \cite{skyrmions5}, which may be interpreted as baryons \cite{nappi1} - \cite{wood2}. 
Here we shall consider a generalized version of the Skyrme model frequently employed in applications of the Skyrme model to nuclear and strong interaction physics \cite{sextic1} - \cite{sextic5}, which is at most quadratic in first time derivatives and, therefore, still has a standard hamiltonian formulation. Its lagrangian density reads
\be \label{gen-Sk-lag}
{\cal L}= \nu_2 {\cal L}_2 + \nu_4 {\cal L}_4 + \nu_0 {\cal L}_0 + \nu_6 {\cal L}_6
\ee
(the $\nu_i$ are some dimensionful coupling constants),
where the terms
\be
{\cal L}_2 = -{\rm tr} (R_\mu R^\mu ) \; , \quad {\cal L}_4 = {\rm tr} ([R_\mu , R_\nu][R^\mu , R^\nu])
\ee
define the Skyrme model originally introduced and studied by Skyrme. Here $R_\mu = (\partial_\mu U)U^\dagger $ is the right-invariant Maurer-Cartan current. The two remaining terms are the potential
\be
{\cal L}_0 = -V(U) 
\ee
which is assumed non-negative and with one unique vacuum in the present paper,
and the sextic term
\be
 {\cal L}_6 = {\cal B}_\mu^2 
\ee
where 
\be
{\cal B}^\mu = -\frac{1}{24\pi^2} \epsilon^{\mu\nu\rho\sigma} {\rm tr} (R_\nu R_\rho R_\sigma )
\ee
is the topological current density giving rise to the integer-valued topological degree
\be
B=\int d^3 x {\cal B}^0 \in \mathbb{Z}
\ee
which may be identified with the baryon number. 

The energy functional for static field configurations of the original Skyrme model (the submodel $\nu_2 {\cal L}_2 + \nu_4 {\cal L}_4$) is known to have a lower bound linear in the topological charge $B$, the so-called Skyrme-Faddeev bound \cite{skyrme1}, \cite{faddeev}. It may be proven easily, however, that nontrivial soliton solutions cannot saturate this bound. One consequence of this is that higher $B$ solitons of the original Skyrme model have rather high binding energies (see, e.g., \cite{manton-book}), which is at odds with the low binding energies of physical nuclei they are supposed to describe. It has been found recently that the submodel $\nu_0 {\cal L}_0 + \nu_6 {\cal L}_6$, too, has a BPS bound linear in $B$ and that, further, nontrivial soliton solutions saturating the bound do exist in this case \cite{BPS-Sk1} - \cite{BPS-Sk3}. This result leads to the proposal to use a version of the Skyrme model for the description of physical nuclei where the  numerical values of the $\nu_i$ parameters are such that the terms ${\cal L}_0$ and ${\cal L}_6$ give the main contributions to the static soliton energies (i.e., nuclear masses). Some first steps in this direction have already been done, with notable success \cite{marleau1} - \cite{BPS-Sk-rop}.

It is the main purpose of the present paper to demonstate that there exist two more topological bounds in the generalized Skyrme model (\ref{gen-Sk-lag}), generalizing (and somewhat simplifying) some recent work by D. Harland on topological energy bounds in the standard Skyrme model with a potential (pion mass) term \cite{harland}. The existence of altogether four bounds implies that each term in the energy functional participates in two bounds and may be distributed arbitrarily among them, which results in a four-parameter family of topological bounds for the full energy. The optimal bound (the optimal values of the four parameters) depend on the specific model, that is, on the form of the potential $V$ and on the values of the coupling constants $\nu_i$.

Our paper is organized as follows. In Section II, we provide the mathematical concepts neede for the derivation of the bounds and then derive the general energy bound for the generalized Skyrme model (\ref{gen-Sk-lag}). In a next step, we consider the bounds for certain submodels, where we discuss in some more detail the cases we believe are of special importance for applications to nuclear physics. We also briefly discuss some generalizations (i.e., bounds for some models which do not belong to the class of generalized Skyrme models (\ref{gen-Sk-lag})). In Section III, we consider the generalized Skyrme model on compact base spaces and derive the additional topological energy bounds which hold in this case. The additional bound grow, in general, faster than linear in the baryon number. These results may be relevant for nuclear matter in the limit of infinite baryon number where, e.g., the formation of crystal-like structures implies the effective compactification of the base space to a torus. In Section IV, we briefly consider the case of Skyrme models in space dimensions different from $d=3$.

\section{Bounds for the Skyrme model}

The tools needed for our calculations of the bounds are just the standard completion of squares in the energy density, the AM-GM (arithmetic mean - geometric mean) inequality
\be
\sum_{i=1}^n a_i \ge n \left( \prod_{i=1}^n a_i \right)^\frac{1}{n} \; , \quad a_i \ge 0
\ee
(with equality iff all $a_i$ are equal),
and the following observation. The group SU(2) as a manifold may be identified with the three-sphere $\mathbb{S}^3$, so static Skyrme fields $U(\vec x)$ are maps ${\bf U}: \mathbb{R}^3 \to \mathbb{S}^3$. Then the topological charge density ${\cal B}_0$ may be naturally extended to a three-form ${\cal B}= {\cal B}_0 d^3 x$, and ${\cal B}$ is just the pullback under the map ${\bf U}$ of the volume three-form $\Omega_{\mathbb{S}^3}$ on $\mathbb{S}^3$, divided by the volume of $\mathbb{S}^3$,
\be
{\cal B} = \frac{1}{{\rm Vol}_{\mathbb{S}^3}}{\bf U}^*(\Omega_{\mathbb{S}^3}) 
\ee
such that
\be
\int_{\mathbb{R}^3} {\cal B} = B \frac{1}{{\rm Vol}_{\mathbb{S}^3}} \int_{\mathbb{S}^3} \Omega_{\mathbb{S}^3} =B
\ee
where the factor $B$ takes into account that the target space is wrapped $B$ times while the base space is covered once. But this implies that when ${\cal B}$ is multiplied by a function of the field variables $f(U)$, it may still be interpreted as the pullback of a target space three-form,
\be
f(U) {\cal B} = \frac{1}{{\rm Vol}_{\mathbb{S}^3}}{\bf U}^*(f(U)\Omega_{\mathbb{S}^3}) 
\ee
with the resulting integral
\be
\int_{\mathbb{R}^3} {f(U)\cal B} = B \frac{1}{{\rm Vol}_{\mathbb{S}^3}} \int_{\mathbb{S}^3} f(U) \Omega_{\mathbb{S}^3} \equiv B\langle f(U)\rangle
\ee
where $\langle f(U)\rangle$ is just the average value of the target space function $f(U)$ when integrated over the whole target space. This result implies that not only the topological charge density ${\cal B}_0$ but also expressions like $f(U){\cal B}_0$ in the energy density are good candidates for topological bounds.

\subsection{Generalized Skyrme model}

There exists a more geometric description of the static energy density of the Skyrme model, originally due to Manton \cite{manton}, which turns out to be extremely useful for our purposes. Using an analogy to elasticity theory, a strain tensor
\be
D_{jk} = -\frac{1}{2}{\rm tr} (R_j R_k )
\ee
may be defined such that all contributions to the static energy density except the one from the potential may be expressed by its eigenvalues.
Indeed, $D_{jk}$ is a symmetric, positive $3\times 3$ matrix with three non-negative eigenvalues $\tilde \lambda_1^2$, $\tilde \lambda_2^2$ and $\tilde \lambda_3^2$. We also define the rescaled (roots of) eigenvalues $\lambda_i = \tilde \lambda_i /\sqrt[3]{2\pi^2}$ which allows to express the topological charge density like
\be
{\cal B}_0 = \frac{1}{2\pi^2} \tilde \lambda_1 \tilde \lambda_2 \tilde \lambda_3 = \lambda_1 \lambda_2 \lambda_3.
\ee
The use of $\lambda_i$ instead of the (roots of the) eigenvalues $\tilde \lambda_i$ avoids factors of $2\pi^2$ and simplifies the expressions below. We now introduce an energy unit $\Lambda$ and a length unit $l$ and measure all energies and lengths in these units, such that our resulting energy expressions and coordinates $\vec x$ are dimensionless. The energy functional of the generalized Skyrme model may then be written like
\begin{equation}
E=\mu_0E_0+\mu_2E_2+\mu_4E_4+\mu_6E_6
\end{equation}
where
\begin{equation}
E_2=\int d^3 x (\lambda^2_1+\lambda^2_2+\lambda^2_3),
\end{equation}
\begin{equation}
E_4=\int d^3 x (\lambda_1^2\lambda_2^2+\lambda_2^2\lambda_3^2+\lambda_3^2\lambda_1^2),
\end{equation}
\begin{equation}
E_6=\int d^3 x \lambda_1^2\lambda_2^2\lambda_3^2 ,
\end{equation}
and
\be
E_0 = \int d^3 x V(U).
\ee
Here the $\mu_i$ are dimensionless coupling constants.
Further, the baryon number (topological degree) is 
\begin{equation}
B = \int d^3 x \lambda_1\lambda_2\lambda_3 .
\end{equation}

Now we want to show that there are altogether four separate topological bounds. In what follows, $\beta$ always is a positive real number. First of all, there is the well-known Skyrme-Faddeev bound
$\beta E_2+E_4 \geq 6 \beta^\frac{1}{2}|B|$,
\begin{eqnarray}
\beta E_2+E_4&=&4 \int d^3 x \frac{1}{4} \left(\beta \lambda_1^2+\beta \lambda_2^2+ \beta \lambda_3^2 \;\;+\;\; \lambda_1^2\lambda_2^2+\lambda_2^2\lambda_3^2+\lambda_3^2\lambda_1^2 \right)  \nonumber \\ 
&=&  \int d^3 x \left( \beta^\frac{1}{2}\lambda_1 \mp \lambda_2\lambda_3 \right)^2+\left( \beta^\frac{1}{2}\lambda_2 \mp \lambda_3\lambda_1 \right)^2+\left( \beta^\frac{1}{2} \lambda_3 \mp \lambda_2\lambda_1 \right)^2  \nonumber \\ && 
\pm  2\cdot 3 \beta^\frac{1}{2} \int d^3x \lambda_1\lambda_2\lambda_3  \quad 
\geq \quad 6 \beta^\frac{1}{2} |B|.
\end{eqnarray}
Then there is the bound of the BPS submodel \cite{BPS-Sk1}, $\beta E_0+E_6 \geq 2 \beta^\frac{1}{2} \langle V^{\frac{1}{2}}\rangle |B|$,
\begin{eqnarray}
\beta E_0+E_6&=&\int d^3 x \left( \lambda_1^2\lambda_2^2\lambda_3^2+\beta V \right) \nonumber \\ & 
=&  \int d^3 x \left( \lambda_1\lambda_2\lambda_3 \mp \sqrt{\beta V}\right)^2  \pm 2 \beta^\frac{1}{2} \int d^3x\lambda_1\lambda_2\lambda_3  \sqrt{V}  \nonumber \\ & 
\geq &   2 \beta^\frac{1}{2} |\int d^3x\lambda_1\lambda_2\lambda_3  \sqrt{V}  | 
= 2 \beta^\frac{1}{2} \langle V^{\frac{1}{2}}\rangle |B|.
\end{eqnarray}
We remark that this bound, and its higher-dimensional generalizations for $E= \beta E_0 + E_{2d}$ in $d$ space dimensions (see below), are special because they can be saturated for arbitrary $B$ and for rather arbitrary (sufficiently well-behaved) potentials, see \cite{BPS-Sk3}, \cite{self-dual}. A further, interesting consequence is that the BPS skyrmion solutions saturating the bound (with positive baryon charge, say) cannot have regions of negative baryon density, which is at variance with the situation in the standard Skyrme model \cite{krusch}.

Further, there exist the following two bounds. The bound $\beta E_0+E_4 \geq 4 \beta^\frac{1}{4} \langle V^{\frac{1}{4}} \rangle |B|$ which was originally found by D. Harland \cite{harland} (using a slightly different derivation involving also a Hoelder inequality),
\begin{eqnarray}
\beta E_0+E_4&=&4 \int d^3 x \frac{1}{4} \left( \lambda_1^2\lambda_2^2+\lambda_2^2\lambda_3^2+\lambda_3^2\lambda_1^2 +\beta V\right) \nonumber \\ & 
\geq &4 \int d^3 x \left( \lambda_1^4\lambda_2^4\lambda_3^4 \beta V\right)^{\frac{1}{4}} 
= 4 \beta^\frac{1}{4} \int d^3 x |\lambda_1\lambda_2\lambda_3| V^{\frac{1}{4}} 
\nonumber \\ & 
\geq & 4 \beta^\frac{1}{4} |\int d^3 x \lambda_1\lambda_2\lambda_3 V^{\frac{1}{4}}| = 4 \beta^\frac{1}{4} \langle V^{\frac{1}{4}}\rangle |B| ,
\label{E04-bound}
\end{eqnarray}
and the bound  $\beta E_2+E_6 \geq 4 \beta^\frac{3}{4} |B|$,
\begin{eqnarray}
\beta E_2+E_6&=&4 \int d^3 x \frac{1}{4} \left(\beta \lambda_1^2+\beta \lambda_2^2+ \beta \lambda_3^2 \;\;+\;\; \lambda_1^2\lambda_2^2\lambda_3^2 \right)  \nonumber \\ & 
\geq &4 \int d^3 x \left( \beta^3 \lambda_1^4\lambda_2^4\lambda_3^4 \right)^{\frac{1}{4}} 
=   4 \beta^\frac{3}{4} \int d^3x|\lambda_1\lambda_2\lambda_3|  \nonumber \\ & 
\geq  & 4| \beta^\frac{3}{4} \int d^3x\lambda_1\lambda_2\lambda_3 |
= 4 \beta^\frac{3}{4} |B|.
\end{eqnarray}
To arrive at the four parameter family of bounds for the full energy we now introduce four parameters $\alpha_i$, $0\le \alpha_i \le 1$, which allow to distribute the four energy terms on their four bounds, and use the above inequalities to arrive at
\begin{eqnarray}
E&=&\mu_0E_0+\mu_2E_2+\mu_4E_4+\mu_6E_6 
\nonumber \\ &=& 
(\mu_0\alpha_0 E_0+ \mu_6\alpha_6 E_6) + (\mu_0 (1-\alpha_0) E_0 + \mu_4 \alpha_4 E_4) + 
\nonumber \\ && ( \mu_2\alpha_2 E_2+\mu_4 (1-\alpha_4)E_4)+(\mu_2(1-\alpha_2)E_2+ \mu_6(1-\alpha_6)E_6)
\nonumber \\ &=& 
\mu_6\alpha_6 \left(\frac{\mu_0\alpha_0}{\mu_6\alpha_6} E_0+  E_6\right) + \mu_4 \alpha_4  \left(\frac{\mu_0 (1-\alpha_0)}{\mu_4 \alpha_4 } E_0 + E_4 \right) + 
\nonumber \\ && \mu_4 (1-\alpha_4) \left( \frac{\mu_2\alpha_2}{\mu_4 (1-\alpha_4)} E_2+E_4 \right)+\mu_6(1-\alpha_6) \left(\frac{\mu_2(1-\alpha_2)}{\mu_6(1-\alpha_6)}E_2+ E_6 \right)   \nonumber \\
 &\geq & 
\left\{ \mu_6\alpha_6 \left(\frac{\mu_0\alpha_0}{\mu_6\alpha_6} \right)^\frac{1}{2}\cdot 2 \langle V^\frac{1}{2} \rangle + \mu_4 \alpha_4  \left(\frac{\mu_0 (1-\alpha_0)}{\mu_4 \alpha_4 }  \right)^\frac{1}{4} \cdot 4  \langle V^\frac{1}{4} \rangle  +  \right.
\nonumber \\ && \left.  6 \mu_4 (1-\alpha_4) \left( \frac{\mu_2\alpha_2}{\mu_4 (1-\alpha_4)}  \right)^\frac{1}{2}+ 4 \mu_6(1-\alpha_6) \left(\frac{\mu_2(1-\alpha_2)}{\mu_6(1-\alpha_6)}\right)^\frac{3}{4}  \right\} |B| 
  \nonumber \\ &=&
\left\{  2\langle V^\frac{1}{2} \rangle \left( \mu_0\alpha_0 \mu_6\alpha_6 \right)^\frac{1}{2} + 4 \langle V^\frac{1}{4} \rangle (\mu_4 \alpha_4)^\frac{3}{4}  \left( \mu_0 (1-\alpha_0) \right)^\frac{1}{4}  +  \right.
\nonumber \\ && \left.  6 \left( \mu_4 (1-\alpha_4) \mu_2\alpha_2 \right)^\frac{1}{2}+ 4 \left( \mu_6(1-\alpha_6) \right)^\frac{1}{4} \left( \mu_2(1-\alpha_2) \right)^\frac{3}{4}  \right\} |B| 
\label{gen-bound}
\end{eqnarray}
which is our main result. The parameter values $\alpha_{2k}$ for the optimal (sharpest) bound obviously depend both on the potential $V$ and on the coupling constants $\mu_i$. We also want to emphasize that the final bound in  (\ref{gen-bound}) is scale invariant, although the initial expression $E$ for the energy is not. Indeed, applying  a scale transformation $\vec x \to \lambda \vec x$ to the Skyrme field, $U_\lambda (\vec x) \equiv U(\lambda \vec x)$, the individual energy terms transform like $E_{2k} \to \lambda^{2k-3}E_{2k}$, which is equivalent to a transformation of the coupling constants $\mu_{2k} \to \lambda^{2k-3}\mu_{2k}$. In the final bound, only scale-invariant combinations like $\mu_6 \mu_0$, $\mu_4^3 \mu_0$, etc., appear.

\subsection{Bounds for submodels}

The exact determination of the optimal parameter values $\alpha_{2k}$ requires the solution of a system of nonlinear algebraic equations which is, in general, not possible analytically.
We may, however, consider one-parameter families of bounds for certain submodels where only three of the four terms contribute, by taking the appropriate limits of the above expression. 

\subsubsection{The model $E_{024}$}

For the model
$E_{024}=\mu_0E_0+\mu_2E_2+\mu_4E_4$ (the standard Skyrme model with a potential), we find the following bound originally derived by D. Harland \cite{harland},
\bea
E&=&\mu_0E_0+\mu_2E_2+\mu_4(1-\alpha_4 + \alpha_4)E_4 \nonumber \\
&\ge& \left( 4\mu_0^\frac{1}{4}(\mu_4(1-\alpha_4))^\frac{3}{4} \langle V^\frac{1}{4} \rangle + 6 (\mu_2\mu_4\alpha_4)^\frac{1}{2} \right) |B|.
\eea
The optimal value for the parameter $\alpha_4$ can be determined by maximizing the bound and reads
\be
\alpha_{4,{\rm opt}} = \frac{1}{2}a^2 \left( \sqrt{1+\frac{4 }{a^2}} - 1\right) 
\; , \quad a^2 \equiv \frac{\mu_2^2}{\mu_0 \mu_4 (\langle V^\frac{1}{4} \rangle )^4 } ,
\ee
leading to the optimal energy bound \cite{harland}
\be \label{E024-bound}
E_{024} \ge \sqrt{\mu_4}|B|\left[ 4\sqrt{\frac{\mu_2}{a}}\left( 1-\frac{a^2}{2}\left( \sqrt{1+\frac{4}{a^2}}-1\right) \right)^\frac{3}{4} + 3\sqrt{2}\sqrt{\mu_2} \; a \left(\sqrt{1+\frac{4}{a^2}}-1\right)^\frac{1}{2}\right] .
\ee
For the standard pion mass potential $V={\rm tr} \; (1-U)$, the author of \cite{harland} compared the soliton energies of numerical solutions for the $B=1$ hedgehog ansatz with the optimal bound (\ref{E024-bound}) for different values of the coupling constants $\mu_{2k}$. For $\mu_0 =0$ (the original Skyrme model without potential) he re-calculated the known result that the hedgehog energy is about 23\% above the Skyrme Faddeev bound, whereas for large $\mu_0$ (equivalently for small $\mu_2$) the hedgehog energy gets closer to the optimal bound and is about 11\% above the bound in the limit $\mu_2 \to 0$. Further, the author of \cite{harland} argued that the situation may become even better (skyrmion energies may get closer to their to their optimal topological bounds) for other choices of potentials. This already indicates that the new topological bounds imply that many generalized Skyrme models lead to soliton energies which are much closer to their topological bounds than previously thought and, consequently, to much smaller binding energies, which is very welcome from a phenomenological point of view. 

\subsubsection{The model $E_{026}$} 
Next, we consider the case $E_{026}=\mu_0E_0+\mu_2E_2+\mu_6E_6$ with the result
\bea
E&=&\mu_0E_0+\mu_2E_2+\mu_6 (1-\alpha_6 + \alpha_6)E_6 \nonumber \\
&\ge& \left( 2 ((\mu_0 \mu_6(1-\alpha_6))^\frac{1}{2} \langle V^\frac{1}{2} \rangle + 4 \mu_2^\frac{3}{4}(\mu_6\alpha_6)^\frac{1}{4} \right) |B|.
\eea
This model is of special interest from the point of view of nuclear physics. On the one hand, it contains the BPS Skyrme model $\mu_0 E_0 + \mu_6 E_6$ as a submodel which, due to its BPS property, is a good starting point for the description of nuclei. On the other hand, it also contains the standard nonlinear sigma model term $E_2$ which, among other features, produces the kinetic energy term for the pion field and is, therefore, required for a more complete and more reliable description of low-energy strong interaction physics. Unfortunately, in this case the maximization of the bound results in a cubic equation for $\alpha_6$  with a rather complicated solution, which we do not display here. 
It is, however, interesting to consider the case when $\mu_2$ is small, i.e.,  when the standard kinetic term $E_2$ is considered as a rather small perturbation of the BPS Skyrme model. For small $\mu_2$ we may perform an expansion in $\mu_2$ or, better, in
\be
b = \frac{\mu_2}{\left( \mu_0^2\mu_6\langle V^\frac{1}{2} \rangle^4 \right)^\frac{1}{3}}
\ee
  which leads to the optimal value for $\alpha_6$
\be
\alpha_{6,{\rm opt}}= b-\frac{2}{3}b^2 +{\bf o}(b^2) .
\ee
The corresponding optimal energy bound, up to linear order in $b$, is
\be
E_{026} \ge 2|B| \sqrt{\mu_0 \mu_6}\langle V^\frac{1}{2}\rangle \left( 1+\frac{3}{2}b\right)  +{\bf o}(b).
\ee
Here the important point is that for $\mu_2 =0$, i.e., for the BPS Skyrme model, skyrmion solutions saturate the bound. Further, for small $\mu_2$ the term $\mu_2 E_2$ will contribute linearly to the skyrmion energy in leading order, i.e.,
\be
E_{026} = 2|B|\sqrt{\mu_0 \mu_6} \langle V^\frac{1}{2} \rangle (1+C_1 b ) + {\bf o}(b) \; , \quad C_1 \ge \frac{3}{2},
\ee
 like in the optimal lower energy bound for $\mu_2 >0$. This optimal bound then implies that for not too large values of $\mu_2$ (i.e., for not too large energy contributions of the term $\mu_2 E_2$), the soliton energies will still be quite close to theri lower bounds, i.e., the model is still "near BPS". The resulting binding energies of higher $B$ skyrmions must, therefore, be small, as is necessary for a reliable application to nuclear physics.

\subsubsection{The model $E_{046}$}
 
The energy functional $E_{046} = \mu_0 E_0+\mu_4 E_4+ \mu_6 E_6 $ has the bound
\bea
E_{046} &=& (1-\alpha_0 + \alpha_0 )\mu_0 E_0 + \mu_4 E_4 + \mu_6 E_6 \nonumber \\
& \geq & \left( 2\langle V^{1/2}\rangle (\mu_0\mu_6
\alpha_0)^{\frac{1}{2}} + 4 \langle V^{\frac{1}{4}} \rangle (\mu_0(1-\alpha_0))^{\frac{1}{4}}
\mu_4^{\frac{3}{4}}  \right) |B| .
\eea
The optimal value for $\alpha_0$ is, again, the solution of a cubic equation which we do not show here. The absence of the term $E_2$ means that this model is probably not adequate for a realistic description of strong interaction physics.

\subsubsection{The model $E_{246}$}

Finally, we may consider the case without potential,
\bea
E&=&\mu_2 (1-\alpha_2 + \alpha_2) E_2+\mu_4 E_4 + \mu_6 E_6 \nonumber \\
&\ge& \left(  6 (\mu_2\mu_4\alpha_2)^\frac{1}{2} +  4 (\mu_2(1-\alpha_2))^\frac{3}{4}(\mu_6)^\frac{1}{4} \right) |B|.
\label{E246}
\eea
In this latter case, the optimal value for $\alpha_2$ reads
\be
\alpha_{2,{\rm opt}} = \frac{1}{2}c^2\; \left( \sqrt{1+\frac{4}{c^2}} - 1\right) \; , \quad
c^2 \equiv \frac{\mu_4^2}{\mu_2 \mu_6},
\ee
and the optimal energy bound is
\be
E_{246}\ge \sqrt{\mu_2}|B| \left[ 3\sqrt{2} \sqrt{\mu_4}\; c \left( \sqrt{1+\frac{4}{c^2}} -1\right)^\frac{1}{2} + 4\sqrt{\frac{\mu_4}{c}}\left( 1-\frac{c^2}{2} \left( \sqrt{1+\frac{4}{c^2}} -1\right) \right)^\frac{3}{4}\right] .
\ee
The model (\ref{E246}) without potential has been studied in \cite{sextic2}, \cite{sextic3} numerically, so let us briefly compare with their results. The authors of \cite{sextic2}, \cite{sextic3} used the energy functional 
\be
\tilde E = -\frac{1}{12\pi^2 }\int d^3 x \left( \frac{1}{2} {\rm tr} R_i^2 +\frac{1-\lambda}{16}{\rm tr} [R_i ,R_j]^2 + 
\frac{\lambda}{96}  {\rm tr} [R_i ,R_j][R_j,R_k][R_k ,R_i] \right)
\ee
(here $\lambda$ is a parameter and $\lambda \in [0,1]$) which, using our rescaled eigenvalues, reads
\bea
\tilde E &=& \frac{1}{12\pi^2 } \int d^3 x \left( \sum_i \tilde \lambda_i^2 + (1-\lambda) \sum_{i<j}\tilde \lambda_i^2 \tilde \lambda_j^2 + \lambda \; \tilde \lambda_1^2 \tilde \lambda_2^2 \tilde \lambda_3^2 \right) \nonumber \\
&=& \frac{\pi^2}{3} \int d^3 x \left( (2\pi^2)^{-\frac{4}{3}} \sum_i \lambda_i^2 + (1-\lambda )
(2\pi^2)^{-\frac{2}{3}}\sum_{i<j} \lambda_i^2 \lambda_j^2 + \lambda \; \lambda_1^2 \lambda_2^2 \lambda_3^2 \right) .
\eea
In \cite{sextic2} explicit numerical values are given for the case $\lambda =1$, i.e., for the submodel without the quartic, Skyrme term, so let us consider this case. The energy and the energy bound are
\be
\tilde E (\lambda =1)= \frac{\pi^2}{3} \left( (2\pi^2)^{-\frac{4}{3}} E_2 + E_6\right)  \ge \frac{2}{3}|B|.
\ee
Numerical energies have been calculated in \cite{sextic2} for baryon numbers $B=1 ,\ldots ,5$. We display these energies per baryon number in Table 1. All energies $\tilde E_B /B$ are above the bound $\tilde E_B /B \ge 2/3$, as must, of course, hold.
\begin{table}
\begin{tabular}{c|ccccc}
$B\quad $&$1$ & 2 & 3 & 4 & 5 \\
\hline 
$\tilde E_B/B \quad$ & $ \; 0.9395 \quad $ & $ 0.864 \quad $ & $ 0.848 \quad $ & 
$ 0.821 \quad $ & $ 0.823 $
\end{tabular}
\caption{}
\end{table}

\subsection{Some further generalizations}

In this paper, we mainly restrict to field theories with lagrangians which are at most quadratic in first time derivatives and, therefore, lead to a standard hamiltonian, but terms not satisfying this constraint have been considered in the literature and may be induced by quantum corrections in an effective field theory, so let us briefly discuss this possibility. The simplest possible term of this type is the standard sigma model term squared, ${\cal L}'_4 = \left( {\rm tr} R_\mu R^\mu )\right)^2 $  leading to the static energy expression
\be
E_4' = \int d^3 x (\lambda_1^2 + \lambda_2^2 + \lambda_3^2 )^2.
\ee
This term may participate in topological energy bounds analogously to the term $E_4$, where just the numerical coefficients are slightly different. Combining it, e.g., with the potential $E_0$ we find the following optimal bound
\bea
E'_4 + \beta E_0 &=& 12 \int d^3 x \frac{1}{12}  \left( \lambda_1^4 + \lambda_2^4 + \lambda_3^4 + \lambda_1^2 \lambda_2^2 + \lambda_1^2 \lambda_2^2 + \lambda_1^2 \lambda_3^2 +
\lambda_1^2 \lambda_3^2 + \lambda_2^3 \lambda_3^2 + \lambda_2^2 \lambda_3^2 
\right. \nonumber \\
&& \left. + 
\frac{\beta}{3} V + \frac{\beta}{3} V + \frac{\beta}{3} V \right) \nonumber \\
&\ge & 12 \int d^3 x \left( \lambda_1^{12} \lambda_2^{12} \lambda_3^{12} \left( \frac{\beta}{3} V\right)^3 \right)^\frac{1}{12} \nonumber \\
&\ge & 4\cdot 3^\frac{3}{4}\beta^\frac{1}{4} |B| \langle V^\frac{1}{4} \rangle .
\label{prime1}
\eea
Further, there exists a (still non-negative) linear combination of $E_4$ and $E'_4$ such that the terms $\lambda_i^2 \lambda_j^2$, $i<j$ terms are absent and the bound is exactly equal to the 
bound for $E_4 + \beta E_0$, namely
\bea
E'_4 - 2 E_4 + \beta E_0 &=& 4\int d^3 x \frac{1}{4}\left( \lambda_1^4 + \lambda_2^4 + \lambda_3^4 + \beta V  \right) \nonumber \\
&\ge & 4 \int d^3 x \left( \lambda_1^4 \lambda_2^4 \lambda_3^4 \beta V \right)^\frac{1}{4} \ge
4\beta^\frac{1}{4} |B| \langle V^\frac{1}{4} \rangle .
\label{prime2}
\eea
Next, let us consider yet another type of field theories. In the literature, sometimes scale-invariant field theories with non-integer powers of kinetic terms have been considered as a way to circumvent the Derrick theorem. The first model of this type, the Deser-Duff-Isham (DDI) model introduced in \cite{deser}, has lagrangian ${\cal L}_3  = {\cal L}_2^\frac{3}{2} =\left( -{\rm tr} R_\mu R^\mu )\right)^\frac{3}{2}$. The root in the lagrangian may lead to problems for general time-dependent configurations, but as far as static configurations are concerned, this model may be treated analogously to the ones considered so far. Indeed, the static energy is scale invariant and, therefore, may support finite energy solutions on its own, and we find
\bea \label{E3}
E_3 & =& \int d^3 x \left(\lambda_1^2+\lambda_2^2+\lambda_3^2)^\frac{3}{2} \right) \ge  3^\frac{3}{2} \int d^3 x
| \lambda_1\lambda_2\lambda_3 | \ge 3^\frac{3}{2}|B|.
\eea
Another possibility consists in taking the Skyrme term to the power $\frac{3}{4}$, ${\cal L}'_3  = {\cal L}_4^\frac{3}{4}$, with the static energy and bound
\bea \label{E3'}
E'_3 &=& \int d^3 x \left(\lambda_1^2\lambda_2^2+\lambda_1^2
\lambda_3^2+\lambda_2^2\lambda_3^2 \right)^\frac{3}{4} \ge  3^\frac{3}{4} \int d^3 x 
| \lambda_1\lambda_2\lambda_3 | \ge 3^\frac{3}{4}|B|.
\eea
We remark that for the analogous models with a target space $\mathbb{S}^2$ (the Nicole \cite{nicole} - \cite{nico2} and the Aratyn-Ferreira-Zimerman   (AFZ) \cite{AFZ1}, \cite{AFZ2} models), the corresponding energy bounds have been found in \cite{harland}.

\subsection{Saturating the bounds}

Finally, let us briefly discuss the possibility of saturating the bounds. As mentioned already, the bound of the BPS Skyrme model $E_6 + \beta E_0$ can be saturated for arbitrary baryon number and for rather arbitrary potentials, whereas the Skyrme-Faddeev bound cannot be saturated (except for the trivial configuration $U=$ const.). For the model $E_4 + \beta E_0$, the possibility to saturate the bound was already discussed in \cite{harland}. Indeed, it follows easily from the derivation (\ref{E04-bound}) that the inequality turns into an equality iff the following conditions are satisfied,
\be \label{sat-cond1}
\lambda_1 = \lambda_2 = \lambda_3 \equiv \lambda
\ee
and
\be \label{sat-cond2}
\beta V(U(x)) = \lambda(x)^4 .
\ee  
Here, condition (\ref{sat-cond1}) is very restrictive. It implies that the strain tensor is proportional to the identity (remember $\tilde \lambda = \sqrt[3]{2\pi^2} \lambda$),
\be
D_{jk} = \tilde\lambda^2 \delta_{jk}
\ee
or, in more geometric terms, that the map ${\bf U}  : \; \mathbb{R}^3 \to \mathbb{S}^3$ induced by the Skyrme field $U$ pulls back the target space metric $ds_{\mathbb{S}^3}^2$ to the Euclidean base space metric, up to a conformal factor, i.e., ${\bf U}^*(ds_{\mathbb{S}^3}^2 ) =
\tilde\lambda^2 (x) ds_{\mathbb{R}^3}^2$. This geometric point of view was used in \cite{harland}. In the same paper it was demonstrated that the inverse stereographic projection from $\mathbb{R}^3$ to $\mathbb{S}^3$ provides a solution with baryon number $B=1$ to condition (\ref{sat-cond1}), where the resulting potential obeying (\ref{sat-cond2}) is 
\be \label{pot-4}
V\sim (1+\phi^0)^4 = \left( \frac{1}{2}{\rm tr} \; (1+U) \right)^4
\ee
 (where $U= \phi^0 + i \sigma^k \phi^k$). Due to the restrictive nature of conditions (\ref{sat-cond1}), this probably is the only solution, although this issue should be further investigated. The fact that the $B=1$ skyrmion (the hedgehog) of the submodel $\mu_0 E_0 + \mu_4 E_4$ with the potential (\ref{pot-4}) saturates the topological energy bound, whereas higher $B$ skyrmions do not saturate it, implies that higher $B$ skyrmions of this submodel are unstable against decay into thier $B=1$ constituents. As was pointed out in \cite{harland}, this implies that an inclusion of the potential (\ref{pot-4}) into the Skyrme model should reduce binding energies, which is again welcome from a phenomenological point of view.

Concerning additional possibilities to saturate energy bounds, we just want to add the observations that the inequality (\ref{prime2}) is saturated by exactly the same field configuration and potential as the inequality (\ref{E04-bound}) (i.e., the inverse stereographic projection and the potential (\ref{pot-4})), whereas (\ref{prime1}) is saturated by the same Skyrme field configuration but with a potential which is three times bigger, i.e., $\beta  V = 3 \lambda^4$. 

Finally, for the models (\ref{E3}) (the DDI model) and (\ref{E3'}), the condition (\ref{sat-cond1}) is sufficient, so both models have the inverse stereographic projection as a solution saturating the bound. For the DDI model, this solution was already found in \cite{deser}.

\section{Skyrme models on compact domains}
There are two main reasons to study skyrmions on compact base spaces. On the one hand, exact solutions may exist on some base spaces which are not available in Euclidean space, and certain geometrical or topological properties of skyrmions may become more transparent. On the other hand, certain skyrmion configurations of physical relevance are equivalent to skyrmions on compact domains. If skyrmions form, e.g., crystal-type structures in the infinite baryon number limit, then these crystals are effecively equivalent to skyrmions on a torus. Concretely,  
we assume that the base space (the domain for static configurations) of the Skyrme model is a compact manifold ${\cal M}$ with volume form $\Omega_{\cal M}$ and finite volume ${\rm Vol}_{\cal M} = \int_{\cal M} \Omega_{\cal M}$, generalizing the results of \cite{harland} for the standard Skyrme model to more general Skyrme models and dimensions.
We remark that a similar bound for the Faddeev-Skyrme model on a compact domain has already been derived in \cite{speight3}.
First of all, all topological bounds derived above continue to hold on compact domains. However, due to the finite volume of the base space one can find more (and sometimes sharper) topological bounds. For these bounds, we need a version of the Hoelder inequality as an additional tool.
The Hoelder inequality reads
\be
 \left( \int_{\cal M} \Omega_{\cal M}   |f|^p \right)^{1/p}  \left( \int_{\cal M} \Omega_{\cal M} |g|^q \right)^{1/q} \geq \int_{\cal M} \Omega_{\cal M} |fg| \; , \quad \frac{1}{p} + \frac{1}{q} =1
\ee
which on a compact space and for $g=1$ gives
$$ \left( \int_{\cal M} \Omega_{\cal M} |f|^p \right)^{1/p}  \left( \int_{\cal M} \Omega_{\cal M} \right)^{1/q} \geq \int_{\cal M} \Omega_{\cal M} |f|$$
or 
\be 
\left( \int_{\cal M} \Omega_{\cal M} |f|^p \right)   \geq \frac{1}{({\rm Vol}_{\cal M})^{p/q}}\left( \int_{\cal M} \Omega_{\cal M} |f| \right)^p .
\ee
We remark that in the cases we shall consider in this section it is only the finite volume of the base space which gives rise to the additional bounds, and, consequently, all bounds are in terms of the baryon number $B$. For base spaces with a nontrivial topology, further topological bounds related to topological invariants of the base space may exist. Specifically, nontrivial bounds may exist even in the sector of field configurations with baryon number zero. A specific example of this possibility has been studied in \cite{canfora}, where the standard Skyrme model was considered on the base space $\mathbb{R}\times \mathbb{T}^2$. There, the author found a nontrivial BPS bound and solitons saturating the bound in the $B=0$ sector, where the nontrivial character of the bound (the nonzero BPS energy) is related to nonzero winding numbers about the two compact directions of the two-torus $\mathbb{T}^2$.

\subsection{Further bounds on compact domains}
In addition to the bounds derived in the previous section for base space $\mathbb{R}^3$, we have the following bounds,
\begin{enumerate}
\item $E_6  \geq \frac{1}{{\rm Vol}_{\cal M}}  |B|^{2}$
\begin{eqnarray}
&E_6&=
\int_{\cal M} \Omega_{\cal M} \lambda_1^2\lambda_2^2\lambda_3^2 = \int_{\cal M} \Omega_{\cal M}  (\mathcal{B}_0)^2
    \nonumber \\ &&
  \geq \frac{1}{{\rm Vol}_{\cal M}}  \left( \int_{\cal M} \Omega_{\cal M} \mathcal{B}_0 \right)^{2} =  \frac{1}{{\rm Vol}_{\cal M}} |B|^{2}
\label{comp-E6-bound}
\end{eqnarray}
\item $E_4 \geq \frac{3}{{\rm Vol}_{\cal M}^\frac{1}{3}} B^\frac{4}{3}$
\begin{eqnarray}
&E_4&=3 \int_{\cal M} \Omega_{\cal M} \frac{1}{3} \left( \lambda_1^2\lambda_2^2+ \lambda_2^2\lambda_3^2 +\lambda_1^2\lambda_3^2 \right)
    \nonumber \\ &&
\geq 3 \int_{\cal M} \Omega_{\cal M} \left( \lambda_1^4\lambda_2^4\lambda_3^4\right)^\frac{1}{3}
    =3 \int_{\cal M} \Omega_{\cal M} |\mathcal{B}_0 |^\frac{4}{3}
        \nonumber \\ &&
    \geq \frac{3}{{\rm Vol}_{\cal M}^\frac{1}{3}} \left( |\int_{\cal M} \Omega_{\cal M} \mathcal{B}_0 | \right)^\frac{4}{3} = \frac{3}{{\rm Vol}_{\cal M}^\frac{1}{3}}  |B|^\frac{4}{3}
\end{eqnarray}
\end{enumerate}
Here, the latter bound (for $E_4$) has already been found in \cite{harland}.
These new bounds may lead to two possible bounds for one and the same model. For instance, for the BPS Skyrme model $\beta E_0 + E_6$ we find the two inequalities
\begin{equation} \label{S3-lin-bound}
\beta E_0+E_6 \geq 2 \beta^\frac{1}{2} \langle V^\frac{1}{2}\rangle  |B| 
\end{equation}
and
\begin{equation} \label{S3-quad-bound}
\beta E_0+E_6 \geq E_6 \geq \frac{1}{{\rm Vol}_{\cal M}} |B|^2 .
\end{equation}
At this point, several comments are in order. Firstly, for sufficiently large $|B|$, the second bound is obviously sharper and soliton energies must grow at least like $|B|^2$. Secondly, while the second bound is, in general, a strict inequality (i.e., nontrivial solutions cannot saturate it), the first bound is, in fact, a BPS bound with BPS soliton solutions saturating it, at least on base space $\mathbb{R}^3$. So the natural question arises whether BPS solutions saturating the linear bound may still exist on compact base spaces, for sufficiently small baryon number $B$. It turns out that under certain circumstances this is, indeed, the case. The BPS solutions on $\mathbb{R}^3$ are of the compacton type for a large class of potentials, i.e., they differ from their vacuum value only on a subspace with finite volume (usually with the topology of a ball or disc). Using the same potentials on ${\cal M}$ and choosing the right values of the coupling constants (or a base space with sufficiently large volume), the resulting soliton solutions are still "compact" solutions of the BPS equations for sufficiently small $|B|$. Here "compact" means that they take non-vacuum values only in a subregion of the full (compact) base space. The size of the "compactons", however, grows with the baryon number $B$, and there exists a certain value $B=B_0$ such that BPS solitons for $|B|\ge B_0$ no longer fit into $\cal{M}$, i.e., formal local solutions of the BPS equations cannot be extended to solutions on the whole base space fulfilling all the required boundary conditions. For these larger values of $B$, solitons are solutions of the full static second-order equations with energies growing at least quadratically in $B$. As the co-existence of both BPS and non-BPS solutions for one and the same model is quite interesting, we shall construct an explicit example displaying this behavior in the next subsection.

\subsection{BPS Skyrmions on $\mathbb{S}^3$}
Here we construct an explicit example of soliton solutions of the BPS Skyrme model on the three-sphere, where the solutions are BPS solutions saturating the bound (\ref{S3-lin-bound}) for sufficiently small baryon number $B$, whereas they are solutions to the full second-order static Euler-Lagrange equations  respecting the second bound (\ref{S3-quad-bound}) for large $B$. It turns out that for the symmetric ansatz we shall use, the second order ODE resulting from the Euler-Lagrange equation can always be integrated once to a first order ODE. The difference between BPS and non-BPS solutions is related to the corresponding integration constant, which is zero for BPS solutions but nonzero for non-BPS solutions.
We use the standard parametrization of the Skyrme model
\be
U= \cos \xi +i \sin \xi \; \vec n \cdot \vec \tau \; , \quad \vec n^2 =1,
\ee
and the stereographic projection
\be
\vec n = \frac{1}{1 + |u|^2} \big( - i (u - \bar u),\, u + \bar u,\, |u|^2-1 \big) ,
\ee
then the BPS Skyrme model $E= \mu_6 E_6 + \mu_0 E_0$ on $\mathbb{S}^3$ may be written as
\begin{equation}
E=\int_{\mathbb{S}^3} \Omega_{\mathbb{S}^3} \left[ \frac{\lambda^2 \sin^4 \xi}{(1+|u|^2)^4} \left( 
i(\nabla \xi )\cdot (\nabla u \times \nabla \bar{u})\right) ^2+\mu^2 V \right]
\end{equation}
where we conveniently introduced new coupling constants $\lambda$ and $\mu$, and $\nabla = \hat e_ a E_a{}^i \partial_i$. Here, $\hat e^a$ is a set of three orthonormal unit vectors, and $E_a{}^i$ is the inverse vielbein which, for a diagonal metric, has only diagonal entries which coincide with the roots of the entries of the inverse metric.
The corresponding BPS equation reads 
\begin{equation}
\frac{\lambda \sin^2 \xi}{(1+|u|^2)^2}  i(\nabla \xi )\cdot (\nabla u \times \nabla \bar{u}) = \pm \mu \sqrt{V} .
\end{equation}
Now, we use the standard metric on a 3-sphere
\begin{equation}
ds_{\mathbb{S}^3}^2=R_0^2(d\rho^2 + \sin^2\rho \; d s_{\mathbb{S}^2}^2 )= R_0^2(d\rho^2 + \sin^2 \rho (d \theta^2 +\sin^2 \theta d\phi^2) )
\end{equation}
where $\rho, \theta \in [0,\pi)$ and $\phi \in [0,2\pi)$. We assume the following ansatz  
\begin{equation}
\xi = \xi (\rho), \;\;\;\ u=u(\theta, \phi)=v(\theta) e^{in\phi}
\end{equation}
where $n$ is equal to the baryon number, $n=B$. This results in
\begin{equation}
v=\tan \frac{\theta}{2},
\end{equation}
and we are left with a first order ODE for the profile function 
\begin{equation}
\frac{2n\lambda \sin^2 \xi}{R_0^3 \sin^2 \rho} \xi_\rho=\pm \mu \sqrt{V(\xi)} .
\end{equation}
It is convenient to introduce a new variable $z=\frac{1}{2}(\rho - \sin \rho \cos \rho)$. Then, $z \in [0,\pi/2)$ and 
\begin{equation}
\frac{2n\lambda \sin^2 \xi}{R_0^3} \xi_z=\pm \mu \sqrt{V(\xi)}  .
\end{equation}
Let us now restrict to the standard Skyrme potential
\begin{equation}
V=1-\cos \xi
\end{equation}
then the profile equation is
\begin{equation}
\frac{2n\lambda \sin^2 \xi}{R_0^3} \xi_z=\pm \mu \sqrt{1-\cos \xi}  
\end{equation}
or
\begin{equation}
\frac{4\sqrt{2} n\lambda}{\mu R_0^3} \cos^2 \frac{\xi}{2} \sin \frac{\xi}{2} d\xi = \pm dz .
\end{equation}
Then,
\begin{equation}
\frac{8\sqrt{2} \lambda n }{3\mu R_0^3} \cos^3 \frac{\xi}{2} = \mp (z-z_0).
\end{equation}
Now we assume the topologically nontrivial boundary conditions (this implies that we have to choose the plus sign)
\begin{equation}
\xi (\rho = 0)=\pi, \;\;\;\; \xi(\rho_0 = \pi)=0
\end{equation}
with $\rho_0 \leq \pi$, which means 
\begin{equation}
\xi (z=0)=\pi, \;\;\;\; \xi(z=z_B)=0
\end{equation}
with $z_B\leq \frac{\pi}{2}$.
Then, solutions are
\begin{equation}
\xi (z)=\left\{ 
\begin{array}{lc}
2 \; \mbox{arccos} \left(\frac{z}{z_B}\right)^\frac{1}{3}  & \quad z \leq z_B \\
 & \\
0 & \quad z \geq z_B
\end{array}
\right. 
\end{equation}
where 
\begin{equation}
z_B=\frac{8\sqrt{2} \lambda}{3\mu R_0^3} \; n
\end{equation}
Obviously, the condition  $z_B\leq \frac{\pi}{2}$ leads to a maximal topological charge which may be carried by such a compact solution,
\begin{equation}
n \leq \frac{3\pi}{16\sqrt{2}} \frac{\mu R_0^3}{\lambda} 
\end{equation}
or 
\begin{equation}
n_{max} =  \left\lfloor \frac{3\pi}{16\sqrt{2}} \frac{\mu R_0^3}{\lambda} \right\rfloor .
\end{equation}
Solitons with a bigger value of the topological charge cannot be solutions of the BPS equation and, therefore, cannot lead to a linear energy-charge relation. The simple reason is that the total volume of such a solution is bigger than the volume of the base space. Then, some of the energy must be used to "squeeze" the solitons. 
\\
These "non-compact" skyrmions are solution of the "generalized BPS equation" (which results from the integration of the second order ODE for the profile function)
\begin{equation}
\frac{2n\lambda \sin^2 \xi}{R_0^3} \xi_z=\pm \mu \sqrt{V(\xi)+V_0}  
\end{equation}
where $V_0$ is an integration constant, which for the standard Skyrme potential reduces to
\begin{equation}
\frac{2n\lambda }{\mu R_0^3} \frac{\sin^2 \xi d\xi}{\sqrt{2\sin^2 \frac{\xi}{2}+V_0}} =dz  .
\end{equation}
Unfortunately, this equation is integrated into a sum of some elliptic functions. Therefore, we will use a different and more suitable potential known as the BPS potential 
\begin{equation}
V(\xi)=\frac{1}{2} (\xi - \cos \xi \sin \xi) .
\end{equation}
Then the general first order equation for the profile function reads
\begin{equation}
\frac{2n\lambda \sin^2 \xi}{R_0^3} \xi_z=\pm \mu \sqrt{\frac{1}{2} (\xi - \cos \xi \sin \xi)+V_0}  .
\end{equation}
It is very convenient to introduce a new target space variable i.e., a new profile function
\begin{equation}
\eta =\frac{1}{2} (\xi - \cos \xi \sin \xi) .
\end{equation}
The last formula can be rewritten as 
\begin{equation}
\frac{2n\lambda}{R_0^3} \eta_z=\pm \mu \sqrt{\eta+\eta_0}  
\end{equation}
with a new integration constant $\eta_0$. 
Further, the boundary conditions are
\begin{equation}
\eta (z=0)= \frac{\pi}{2}, \;\;\; \eta (z=z_B)=0, \;\;\; \mbox{where} \;\;\; z_B \leq \frac{\pi}{2} .
\end{equation}
For the BPS sector we assume $\eta_0=0$. Then the solution is  
\begin{equation}
\eta (z)=\left\{ 
\begin{array}{lc}
\frac{\pi}{2} \left(1- \frac{z}{z_B} \right)^2 & z \leq z_B \\
 & \\
0 & z \geq z_B
\end{array}
\right. 
\end{equation}
where 
\begin{equation}
z_B=\frac{2\sqrt{2\pi} \lambda}{\mu R_0^3} \; n .
\end{equation}
Obviously, such a solution makes a sense only if $z_B \leq \pi/2$. Hence, 
\begin{equation}
\frac{2\sqrt{2\pi} \lambda}{\mu R_0^3} \; n \leq \frac{\pi}{2} \;\; \Rightarrow \;\; n \leq \frac{\mu R_0^3}{\lambda} \frac{1}{4} \sqrt{\frac{\pi}{2}} .
\end{equation}
Then, again, solutions do not fill the whole base space completely and the energy is linear with the topological charge. For higher charges we have to consider a non-zero value for $\eta_0$, and the solution is
\begin{equation}
\eta (z) = \frac{1}{\beta^2} \left[ \left( \frac{\beta^2}{2} +\frac{\pi}{4} -z \right)^2 -  \left( \frac{\beta^2}{2} +\frac{\pi}{4}  \right)^2 + \frac{\beta^2 \pi}{2} \right]
\end{equation}
where 
\begin{equation}
\beta = \frac{4\lambda}{\mu R_0^3} n .
\end{equation}
This solution makes sense if
\begin{equation}
\beta^2 \geq \frac{\pi}{2} \;\;\; \Rightarrow \;\;\;  n \geq \frac{\mu R_0^3}{\lambda} \frac{1}{4} \sqrt{\frac{\pi}{2}} .
\end{equation}
The total energy is
\begin{equation}
E=4\pi \mu^2 R_0^3 \left[ \frac{2}{3\beta^2} \left( \frac{\beta^2}{2} +\frac{\pi}{2} \right)^3 -\frac{2}{3\beta^2} \left( \frac{\beta^2}{2} \right)^3 + \frac{\pi}{2} \left(\frac{\pi}{2} -  \frac{1}{\beta^2} \left( \frac{\beta^2}{2} +\frac{\pi}{4} \right)^2 \right) \right] .
\end{equation}
As $\beta \sim n$, then at leading order in $n$ we get
\begin{equation}
E=4\pi^2 \frac{\lambda \mu}{R_0^3}n^2 + O(n^0)
\end{equation}
which confirms the topological bound we found above.

\section{Skyrme models in different dimensions}

From a physical point of view, the Skyrme model and its generalizations in $d=3$ space dimensions are the most relevant ones, but Skyrme models in different dimensions have been studied and are of some interest. The baby Skyrme model in $d=2$ space dimensions \cite{old1} - \cite{foster} has been studied quite intensely, both as a toy model for the full Skyrme model and because it has some independent applications, mainly in condensed matter physics (see,e.g., \cite{qhe1}, \cite{qhe2}). The investigation of Skyrme models in higher dimensions has not been developed thus far until now, although they, too,  may be of some interest, e.g., in the context of brane cosmology \cite{brane1} - \cite{brane3}. The central idea of brane cosmology is that our $3+1$ dimensional universe is a topological defect within a higher-dimensional bulk universe \cite{akama}, \cite{rubakov}. This idea gained momentum when it was found that topological defect solutions of the higher- dimensional Einstein-matter system may exist such that both gravitational and non-gravitational interactions are effectively confined to the topological defect (the brane) \cite{sundrum}, \cite{csaki}. In the simplest setting, the brane is a co-dimension one defect (a domain wall) in a $4+1$ dimensional bulk universe, but branes which are co-dimension $d$ defects in a $(3+d)+1$ dimensional bulk universe are prefectly viable. Skyrmions in $d$ space dimensions provide specific examples of such co-dimension $d$ topological defects, and topological energy bounds for these models may therefore be useful, and we shall briefly discuss them here.
 
The Skyrme model may be viewed as a field theory in three space dimensions with fields taking values in the (target space) three-sphere, and it is this point of view which we want to generalize. So, a Skyrme model in $d$ space plus one time dimensions is a field theory with fields taking values in the d-sphere $\mathbb{S}^d$ described by a unit vector wit $d+1$ components,
\be
\phi^a, \quad \phi^a \phi^a =1, \quad a = 1, \ldots ,d+1.
\ee
If the fields are constrained to take a unique value (e.g., the vacuum value of the potential) at spatial infinity, which we assume, then the field configurations fall into different homotopy classes characterized by an integer topological degree (winding number).  We still restrict to lagrangians which are at most quadratic in time derivatives, then in addition to a potential ${\cal L}_0 = -V(\phi^a)$ we may have the following derivative terms,
\be
{\cal L}_{2k} =  \epsilon^{a_1 \ldots a_k c_{k+1} \ldots c_{d+1}}\epsilon^{\mu_1 \ldots \mu_k \rho_{k+1} \ldots \rho_d}
\phi^{a_1}_{\mu_1} \ldots \phi^{a_k}_{\mu_k} 
\epsilon^{b_1 \ldots b_k c_{k+1} \ldots c_{d+1}}\epsilon^{\nu_1 \ldots \nu_k}{}_{ \rho_{k+1} \ldots \rho_d}
\phi^{b_1}_{\nu_1} \ldots \phi^{b_k}_{\nu_k} 
\ee
where $\phi^a_\mu \equiv \partial_\mu \phi^a$, and $k=1 , \ldots ,d$. 
Finally, the topological degree is
\be
B= \frac{1}{d!{\rm Vol}_{\mathbb{S}^d}}\int d^d x 
\epsilon^{a_0a_1 \ldots a_d}\epsilon^{\mu_1 \ldots \mu_d}\phi^{a_0}\phi_{\mu_1}^{a_1} \ldots \phi_{\mu_d}^{a_d}. 
\ee
We may again define a $d\times d$ "strain tensor" for static field configurations,
\be
D^l_{k} = g^{lj}\phi^a_j \phi^a_k,
\ee
such that all contributions to the static energy (except for the potential) may be expressed in terms of the $d$ non-negative eigenvalues $\tilde \lambda_j^2$ of the strain tensor (here the rising of one index by the inverse base space metric $g^{lj}$ is immaterial for an Euclidean base space, but becomes relevant on general base spaces). Further, we shall again rescale the (roots of the) eigenvalues $\tilde \lambda_j \to \lambda_j$ by a common constant factor such that the topological index is just the integral of the product of all $\lambda_j$,
\be
B = \int d^d x \lambda_1 \ldots \lambda_d \in \mathbb{Z} .
\ee
This avoids clumsy factors in the expressions below.

\subsection{The baby Skyrme model in two space dimensions}
The bound of the baby Skyrme model is, in principle, wellknown, so we just briefly repeat it here. The dimensionless static energy functional reads
\be
E= \mu_0 E_0 + \mu_2 E_2 + \mu_4 E_4 
\ee
where $E_0 = \int d^2 x V$, $E_2 = \int d^2 x (\lambda_1^2 + \lambda_2^2)$ and $E_4 = \int d^2 x \lambda_1^2 \lambda_2^2$. There are two topological bounds, namely \cite{poly}
\be
E_2 = \int d^2 x (\lambda_1 \mp \lambda_2 )^2 \pm 2\int d^2 x \lambda_1 \lambda_2 \ge 2|\int d^2 x \lambda_1 \lambda_2 |= 2 |B|
\ee
and \cite{ward}, \cite{restr-bS}, \cite{Sp1}
\bea\beta E_0 + E_4 &=& \int d^2 x (\beta V + \lambda_1^2 \lambda_2^2 ) = \int d^2 x (\lambda_1 \lambda_ 2 \mp \sqrt{\beta V})^2 \pm 2\sqrt{\beta} \int d^2 x \lambda_1 \lambda_2 \sqrt{V} \nonumber \\ 
&\ge & 2\beta^\frac{1}{2} |\int d^2 x \lambda_1 \lambda_2 V^\frac{1}{2}  | = 2\beta^\frac{1}{2}\langle V^\frac{1}{2} \rangle |B|.
\eea
The total energy is, therefore, bound by
\be
E = \mu_2 E_2 + \mu_4 \left( \frac{\mu_0}{\mu_4} E_0 + E_4 \right)  \ge 2 \left( \mu_2 + (\mu_0 \mu_4)^\frac{1}{2} \langle V^\frac{1}{2} \rangle \right) |B|.
\ee  
On a compact two-dimensional base space ${\cal M}$ there exists the additional bound
\be
E_4 \ge \frac{1}{{\rm Vol}_{\cal M}}|B|^2
\ee
where ${\rm Vol}_{\cal M}$ is the area of the base space. This bound is equivalent to the bound (\ref{comp-E6-bound}) for $E_6$ on a three-dimensional compact base space.

\subsection{Skyrme models in higher dimensions}
The dimensionless static energy of the generalized Skyrme model in 4 dimensions reads
\begin{equation}
E=\mu_0E_0+\mu_2E_2+\mu_4E_4+\mu_6E_6+\mu_8E_8,
\end{equation}
where $E_0 = \int d^d x V$ and, using the rescaled eigenvalues of the strain tensor, the further expressions read
\begin{equation}
E_2=\int d^4 x (\lambda^2_1+\lambda^2_2+\lambda^2_3+\lambda_4^2)
\end{equation}
\begin{equation}
E_4=\int d^4 x (\lambda_1^2\lambda_2^2+\lambda_2^2\lambda_3^2+\lambda_3^2\lambda_1^2+\lambda_1^2\lambda_4^2+\lambda_2^2\lambda_4^2
+\lambda_3^2\lambda_4^2)
\end{equation}
\begin{equation}
E_6=\int d^4 x (\lambda_1^2\lambda_2^2\lambda_3^2+ \lambda_1^2\lambda_2^2\lambda_4^2+ \lambda_1^2\lambda_3^2\lambda_4^2+ \lambda_2^2\lambda_3^2\lambda_4^2)
\end{equation}
\begin{equation}
E_8=\int d^4 x \lambda_1^2\lambda_2^2\lambda_3^2\lambda_4^2
\end{equation}
Further 
\begin{equation}
B = \int d^3 x \lambda_1\lambda_2\lambda_3\lambda_4
\end{equation}
is the pertinent topological charge. Generalizations to 5 or higher dimensions are obvious. Again, in Euclidean space, there exists a separate topological energy bound for each pair of energy expressions $E_{2k}$ which behave oppositely under Derrick scaling, now in four dimensions. In addition, in this case (like in $d=2$ dimensions for $E_2$) there exists a separate bound for the scale invariant term $E_4$. Further, there exist two more bounds (for $E_6$ and $E_8$) on compact base spaces. The calculations of these bounds are similar to the calculations in two and three dimensions and sometimes lead to rather lengthy expressions, therefore we relegate them to the appendix. 

\section{Conclusions}
In the present paper, we investigated the issue of topological energy bounds in generalized Skyrme models, generalizing and complementing the recent results of \cite{harland}. We mostly restricted our considerations to generalizations of the Skyrme model which still lead to a standard hamiltonian. The principal result of this investigation is that in generalized Skyrme models there exists a rather large number of new topological energy bounds which have not been known until now. 
 Apart from being an interesting and unexpected mathematical result on its own, 
the main physical importance of these new bounds liles in the fact that the soliton solutions of generalized Skyrme models obey much sharper energy bounds than thought previously. In other words, quite many generalized Skyrme models are "near-BPS" models with soliton energies rather close to the new, sharper topological bounds, which implies that possible binding energies must be small, exactly as required for an application to nuclear physics. Indeed, if the energy of the $B=1$ skyrmion (the hedgehog) of a generalized Skyrme model is $1+\delta$ times the optimal topological bound, i.e., $E_{B=1} = (1+\delta)E^{(0)}$, where the optimal bound is $E_B\ge |B|E^{(0)}$, then the binding energies $\Delta_B$ of higher skyrmions (the energetic cost of a desintegration into their $B=1$ constituents) are bound by
\be
\Delta_B \equiv |B| E_{B=1} -E_B \le |B| E^{(0)} (1+\delta) - |B| E^{(0)} = |B|E^{(0)} \delta
\ee
and the relative binding energies are bound by $\delta$,
\be
\frac{\Delta_B}{E_B} \le \frac{\Delta_B}{|B|E^{(0)}} \le \delta .
\ee
and are necessarily small for small $\delta$. So the new, sharper bounds significantly extend the space of physically viable generalizations of the Skyrme model. We remark that a different proposal for a near-BPS Skyrme model via the inclusion of vector mesons and based on an instanton holonomy, has been developed recently in \cite{sutcliffe1} - \cite{ma2}.

From a more numerical point of view,    
these new energy bounds may serve as benchmarks for numerical calculations of solitons and soliton energies in generalized Skyrme models, which should be instrumental in a more precise determination of these soliton solutions and in a better control of possible numerical errors or finite-size effects.  

We also considered Skyrme models on compact base spaces, which 
 typically lead to further, additional topological energy bounds related to the finite volume of the base space. 
Skyrme models on compact spaces are relevant for the analysis of very high density
nuclear matter, where skyrmions form crystal-like structures \cite{manton}, \cite{cristal1} - \cite{speight2}, so our new bounds may be helpful in this context. We also found new topological energy bounds for some further nonlinear field theories supporting topological solitons like, e.g., the DDI model \cite{deser}. Finally, we briefly discussed the resulting topological energy bounds for Skyrme models in higher dimensions.

At last, let us mention that in the very recent publication by D. Harland \cite{harland}, in addition to the bounds for the standard Skyrme model with a potential, new energy bounds for the Skyrme-Faddeev model with a potential and for the Nicole and Aratyn-Ferreira-Zimerman models have been found.    

\section*{Acknowledgement}
The authors acknowledge financial support from the Ministry of Education, Culture and Sports, Spain (grant FPA2008-01177), 
the Xunta de Galicia (grant INCITE09.296.035PR, grant GRC2013-024, and
Conselleria de Educacion), the
Spanish Consolider-Ingenio 2010 Programme CPAN (CSD2007-00042), and FEDER. 
Further, AW was supported by polish NCN (National Science Center)
grant DEC-2011/01/B/ST2/00464 (2011-2014).
The authors thank M. Speight for helpful discussions and D. Harland for helpful comments and for informing them on his results prior to publication.

\section*{Appendix}
Here we present some of the calculations and all results for pairwise or individual bounds for general Skyrme models in 4 and 5 dimensions. The generalization to higher dimensions is straight-forward.

\subsection{4 dimensions}

In $d=4$ space dimensions in Euclidean space, there are five separate topological bounds.
\begin{enumerate}
\item $\beta E_0+E_6 \geq 6\cdot 2^{-\frac{1}{3}} \beta^\frac{1}{3} \langle V^{\frac{1}{3}}\rangle |B|$
\begin{eqnarray}
&\beta E_0+E_6&=4 \int d^4x \frac{1}{4} \left(\lambda_1^2\lambda_2^2\lambda_3^2+ \lambda_1^2\lambda_2^2\lambda_4^2+ \lambda_1^2\lambda_3^2\lambda_4^2+ \lambda_2^2\lambda_3^2\lambda_4^2 +\beta V\right) \nonumber \\ && 
= 6 \int d^4 x \frac{1}{6} \left(\lambda_1^2\lambda_2^2\lambda_3^2+ \lambda_1^2\lambda_2^2\lambda_4^2+ \lambda_1^2\lambda_3^2\lambda_4^2+ \lambda_2^2\lambda_3^2\lambda_4^2 +\frac{\beta V}{2}+\frac{\beta V}{2} \right)  \nonumber \\ && 
\ge 6 \int d^4 x \left|\beta^2 \lambda_1^6\lambda_2^6\lambda_3^6\lambda_4^6 \frac{V^2}{4}\right|^{\frac{1}{6}} 
\nonumber \\ && 
= 6\cdot 2^{-\frac{1}{3}} \beta^\frac{1}{3} \int d^4 x | \lambda_1\lambda_2\lambda_3 \lambda_4| V^{\frac{1}{3}} \nonumber \\ && 
\geq  6\cdot 2^{-\frac{1}{3}} \beta^\frac{1}{3} \langle V^{\frac{1}{3}}\rangle |B|
\end{eqnarray}
\item $\beta E_0+E_8 \geq 2  \beta^\frac{1}{2}\langle V^{\frac{1}{2}}\rangle |B|$
\begin{eqnarray}
&\beta E_0+E_8&=\int d^4 x (\lambda_1^2\lambda_2^2\lambda_3^2\lambda_4^2+\beta V)  \nonumber \\ && 
=  \int d^4 x \left( \lambda_1\lambda_2\lambda_3\lambda_4 \mp \sqrt{\beta V}\right)^2 \pm 2\beta^\frac{1}{2}\int d^4x\lambda_1\lambda_2\lambda_3 \lambda_4  \sqrt{V}  \nonumber \\ && 
\geq  2 \beta^\frac{1}{2} |\int d^4x\lambda_1\lambda_2\lambda_3\lambda_4  \sqrt{V}  |  \nonumber \\ && 
= 2 \beta^\frac{1}{2}  \langle V^{\frac{1}{2}}\rangle |B|
\end{eqnarray}
\item $\beta E_2+E_6 \geq 8 \beta^\frac{1}{2} |B|$
\begin{eqnarray}
\beta E_2+E_6&=&4 \int d^4 x \frac{1}{4} \left(\beta (\lambda_1^2+\lambda_2^2+ \lambda_3^2 +\lambda_4^2) +\lambda_1^2\lambda_2^2\lambda_3^2+ \lambda_1^2\lambda_2^2\lambda_4^2+ \lambda_1^2\lambda_3^2\lambda_4^2+ \lambda_2^2\lambda_3^2\lambda_4^2 \right)  \nonumber \\ & 
=& \int d^4 x \left(\beta^\frac{1}{2} \lambda_1 \mp  \lambda_2\lambda_3\lambda_4 \right)^2+ \left(\beta^\frac{1}{2}\lambda_2 \mp  \lambda_1\lambda_3\lambda_4 \right)^2+\left(\beta^\frac{1}{2} \lambda_3 \mp  \lambda_1\lambda_3\lambda_4 \right)^2+
\nonumber \\ && +
\left(\beta^\frac{1}{2}\lambda_4 \mp  \lambda_2\lambda_3\lambda_1 \right)^2
  \pm 4\cdot 2 \beta^\frac{1}{2} \int d^4 x \lambda_1 \lambda_2\lambda_3\lambda_4 
  \nonumber \\ &
\geq  &8 \beta^\frac{1}{2} | \int d^4 x \lambda_1 \lambda_2\lambda_3\lambda_4 |
= 8\beta^\frac{1}{2}|B|
\end{eqnarray}
\item $\beta E_2+E_8 \geq 6 \cdot 2^{-\frac{1}{3}}\beta^\frac{2}{3}|B|$
\begin{eqnarray}
\beta E_2+E_8&=& \int d^4 x \left( \beta (\lambda_1^2+\lambda_2^2+ \lambda_3^2 +\lambda^2_4)+ \lambda_1^2\lambda_2^2\lambda_3^2\lambda_4^2 \right)  \nonumber \\  
&=&6\int d^4 x \frac{1}{6} \left(\beta (\lambda_1^2+\lambda_2^2+ \lambda_3^2 +\lambda_4^2)+ \frac{1}{2}\lambda_1^2\lambda_2^2\lambda_3^2\lambda_4^2+\frac{1}{2}\lambda_1^2\lambda_2^2\lambda_3^2\lambda_4^2 \right)  \nonumber \\  
& \geq & 6 \int d^4 x \left| \beta^4\lambda_1^6\lambda_2^6\lambda_3^6\lambda_4^6 \frac{1}{4}\right|^{1/6}
= 6 \cdot 2^{-\frac{1}{3}} \beta^\frac{2}{3} \int d^4 x \left| \lambda_1\lambda_2\lambda_3\lambda_4 \right|   \nonumber \\ 
& \geq & 6 \cdot 2^{-\frac{1}{3}} \beta^\frac{2}{3} |B|
\end{eqnarray}
\item $E_4 \geq 8 |B|$
\begin{eqnarray}
E_4&=& \int d^4 x (\lambda_1\lambda_2 \mp \lambda_3\lambda_4)^2+(\lambda_1\lambda_3 \mp \lambda_2\lambda_4)^2+(\lambda_1\lambda_4 \mp \lambda_2\lambda_4)^2   \nonumber \\ && 
\pm 4\cdot 2 \int d^4 x \lambda_1\lambda_2\lambda_3\lambda_4
 \quad
\geq  \quad 8 |B|
\end{eqnarray}
\end{enumerate}
On a compact four-dimensional manifold ${\cal M}$ we have, in addition, the following bounds
\begin{enumerate}
\item $E_6  \geq \frac{4}{\sqrt{{\rm Vol}_{\cal M}}}  |B|^\frac{3}{2}$
\begin{eqnarray}
&E_6&=
\int \Omega_{\cal M} \left( \lambda_1^2\lambda_2^2\lambda_3^2+ \lambda_1^2\lambda_2^2\lambda_4^2+ \lambda_1^2\lambda_3^2\lambda_4^2+ \lambda_2^2\lambda_3^2\lambda_4^2 \right) 
  \nonumber \\ &&
  =  4\int  \Omega_{\cal M} \frac{1}{4} \left( \lambda_1^2\lambda_2^2\lambda_3^2+ \lambda_1^2\lambda_2^2\lambda_4^2+ \lambda_1^2\lambda_3^2\lambda_4^2+ \lambda_2^2\lambda_3^2\lambda_4^2 \right)
  \nonumber \\ &&
  \geq 4 \int \Omega_{\cal M}  (\lambda_1^6\lambda_2^6\lambda_3^6\lambda^6_4)^\frac{1}{4} = 4 \int  \Omega_{\cal M} \mathcal{B}^\frac{3}{2}
    \nonumber \\ &&
  \geq \frac{4}{\sqrt{{\rm Vol}_{\cal M}}}  \left( \int \Omega_{\cal M} \mathcal{B} \right)^\frac{3}{2} = \frac{4}{\sqrt{{\rm Vol}_{\cal M}}} |B|^\frac{3}{2}
\end{eqnarray}
\item $E_8 \geq \frac{1}{{\rm Vol}_{\cal M}} B^2$
\begin{eqnarray}
&E_8&=\int \Omega_{\cal M} \lambda_1^2\lambda_2^2\lambda_3^2\lambda_4^2
=\int \Omega_{\cal M} \mathcal{B}^2
    \nonumber \\ &&
    \geq \frac{1}{{\rm Vol}_{\cal M}} \left( \int \Omega_{\cal M} \mathcal{B} \right)^2 = \frac{1}{{\rm Vol}_{\cal M}}  B^2 .
\end{eqnarray}
\end{enumerate}

\subsection{5 dimensions}
The static energy $E$ and topological charge $B$ in 5 dimensions, expressed in terms of the rescaled eigenvalues of the strain tensor, are 
\begin{equation}
E=\mu_0E_0+\mu_2E_2+\mu_4E_4+\mu_6E_6+\mu_8E_8+ \mu_{10}E_{10}
\end{equation}
where
\begin{equation}
E_2=\int d^4 x (\lambda^2_1+\lambda^2_2+\lambda^2_3+\lambda_4^2+\lambda_5^2)
\end{equation}
\begin{equation}
E_4=\int d^4 x (\lambda_1^2\lambda_2^2+\lambda_1^2\lambda_3^2+\lambda_1^2\lambda_4^2+\lambda_1^2\lambda_5^2+\lambda_2^2\lambda_3^2+\lambda_2^2\lambda_4^2+\lambda_2^2\lambda_5^2+\lambda_3^2\lambda_4^2+\lambda_3^2\lambda_5^2+\lambda_4^2\lambda_5^2)
\end{equation}
\begin{eqnarray}
E_6&=&\int d^4 x (\lambda_1^2\lambda_2^2\lambda_3^2+ \lambda_1^2\lambda_2^2\lambda_4^2+ \lambda_1^2\lambda_2^2\lambda_5^2+ \lambda_1^2\lambda_3^2\lambda_4^2+ \lambda_1^2\lambda_3^2\lambda_5^2+ 
  \nonumber \\ && 
  \lambda_1^2\lambda_4^2\lambda_5^2+ \lambda_2^2\lambda_3^2\lambda_4^2+ \lambda_2^2\lambda_3^2\lambda_5^2+ \lambda_2^2\lambda_4^2\lambda_5^2+ \lambda_3^2\lambda_4^2\lambda_5^2)
\end{eqnarray}
\begin{equation}
E_8=\int d^4 x (\lambda_1^2\lambda_2^2\lambda_3^2\lambda_4^2 + \lambda_1^2\lambda_2^2\lambda_3^2\lambda_5^2+ \lambda_1^2\lambda_2^2\lambda_4^2\lambda_5^2+ \lambda_1^2\lambda_3^2\lambda_4^2\lambda_5^2+ \lambda_2^2\lambda_3^2\lambda_4^2\lambda_5^2)
\end{equation}
\begin{equation}
E_{10} = \int d^4 x \lambda_1^2\lambda_2^2\lambda_3^2\lambda_4^2\lambda_5^2
\end{equation}
and
\begin{equation}
B = \int d^3 x \lambda_1\lambda_2\lambda_3\lambda_4\lambda_5.
\end{equation}
In Euclidean space, there are nine separate topological bounds.
\begin{enumerate}
\item $\beta E_0+E_{10} \geq 2 \beta^\frac{1}{2} \langle V^{\frac{1}{2}}\rangle |B|$
\item $\beta E_0+E_8 \geq \frac{8}{3^\frac{3}{8}} \beta^\frac{3}{8}  \langle V^\frac{3}{8}\rangle |B|$
\item $\beta E_0+E_6 \geq \frac{12}{2^\frac{1}{6}}\beta^\frac{1}{6}\langle V^\frac{1}{6}\rangle |B|$
\item $\beta E_2+E_{10} \geq \frac{8}{3^\frac{3}{8}} \beta^\frac{5}{8} |B|$
\item $\beta E_2+E_8 \geq 10 \beta^\frac{1}{2} |B|$
\item $\beta E_2+E_6 \geq \frac{40}{3^\frac{3}{4}2^\frac{1}{4}} \beta^\frac{1}{4} |B|$
\item $\beta E_4+E_{10} \geq \frac{12}{2^\frac{1}{6}} \beta^\frac{5}{6} |B|$
\item $\beta E_4+E_8 \geq \frac{40}{3^\frac{3}{4}2^\frac{1}{4}} \beta^\frac{3}{4}|B|$
\item $\beta E_4 + E_6 \geq 20 \beta^\frac{1}{2} |B|$. 
\end{enumerate}
On a compact base space, there are three additional topological bounds,
\begin{enumerate}
\item $E_{10}  \geq \frac{1}{{\rm Vol}_{\cal M}}  |B|^2$
\item $E_8 \geq \frac{5}{({\rm Vol}_{\cal M})^\frac{3}{5}} B^\frac{8}{5}$
\item $E_6 \geq \frac{10}{({\rm Vol}_{\cal M})^\frac{1}{5}} B^\frac{6}{5}$ .
\end{enumerate}


\begin{thebibliography}{99}

\bibitem{skyrme1} T. H. R. Skyrme, Proc. Roy. Soc. Lon. {\bf 260},
127 (1961).
\bibitem{skyrme2}
T. H. R. Skyrme, Nucl. Phys. {\bf 31}, 556 (1962).
\bibitem{skyrme3}
T. H. R. Skyrme,
 J. Math. Phys. {\bf 12}, 1735 (1971).
\bibitem{skyrmions1}
C. J. Houghton, N. S. Manton, P. M. Sutcliffe,
Nucl. Phys. B{\bf 510}, 507 (1998).
\bibitem{skyrmions2}
R. A. Battye, P. M. Sutcliffe, Nucl. Phys. B{\bf 705}, 384 
(2005).
\bibitem{skyrmions3}
 R. A. Battye, P. M. Sutcliffe, Phys. Rev. C{\bf 73},  055205 (2006).
\bibitem{skyrmions4}
R. A. Battye, N. S. Manton, P. M. Sutcliffe,
Proc. Roy. Soc. Lond. A{\bf 463}, 261 (2007).
\bibitem{skyrmions5}
D. T. J. Feist, P. H. C. Lau, N. S. Manton, Phys. Rev. D{\bf 87}, 085034 (2013).
\bibitem{nappi1} G. S. Adkins, C. R. Nappi, E. Witten, Nucl. Phys. 
B{\bf 228}, 552 (1983).
\bibitem{nappi2} 
G. S. Adkins, C. R. Nappi, Nucl. Phys. B{\bf 233}, 109 
(1984).
\bibitem{braaten1}
E. Braaten, L. Carson,
Phys. Rev. Lett. {\bf 56}, 1897 (1986).
\bibitem{braaten2}
E. Braaten, L. Carson,
Phys. Rev. D{\bf 38}, 3525 (1988).
\bibitem{carson1}
L. Carson,
Phys. Rev. Lett. {\bf 66}, 1406 (1991).
\bibitem{carson2}
L. Carson,
Nucl. Phys. A{\bf 535}, 479 (1991).
\bibitem{walhout}
T. S. Walhout,
Nucl. Phys. A{\bf 531}, 596 (1991). 
\bibitem{wood1}
O. V. Manko, N. S. Manton, S. W. Wood,
Phys. Rev. C{\bf 76}, 055203 (2007).
\bibitem{wood2}
R. A. Battye, N. S. Manton, P. M. Sutcliffe, 
S. W. Wood, Phys. Rev. C{\bf 80}, 034323 (2009).
\bibitem{sextic1} A. Jackson, A. D. Jackson, A. S. Goldhaber,
G. E. Brown, L. C. Castillejo, Phys. Lett. B{\bf 154}, 101 (1985).
\bibitem{sextic2}
L. Floratos, B. M. A. G. Piette, Phys. Rev. {\bf D64}, 045009 (2001).
\bibitem{sextic3}  
L. Floratos, B. M. A. G. Piette,
J. Math. Phys. {\bf 42}, 5580 (2001).
\bibitem{sextic4}
V. B. Kopeliovich, A. M. Shunderuk, G. K. Mathusko, Phys. Atom. Nucl. 69 
(2006) 120.
\bibitem{sextic5}
V. B. Kopeliovich, Phys. Part. Nucl. {\bf 37}, 623 (2006).
\bibitem{faddeev}
L. D. Faddeev, Lett. Math. Phys. {\bf 1}, 289 (1976).
\bibitem{manton-book}
N. Manton, P. Sutcliffe, "Topological Solitons", Cambridge University Press, Cambridge, 2007.
\bibitem{BPS-Sk1}
C. Adam, J. Sanchez-Guillen, A. Wereszczynski,
Phys. Lett. B{\bf 691}, 105 (2010).
\bibitem{BPS-Sk2}
C. Adam, J. Sanchez-Guillen, A. Wereszczynski,
Phys. Rev. D{\bf 82}, 085015 (2010).
\bibitem{BPS-Sk3}
C. Adam, C. D. Fosco, J. M. Queiruga, J. Sanchez-Guillen, A. Wereszczynski,
 J.Phys. A{\bf 46}, 135401 (2013).
\bibitem{marleau1}
E. Bonenfant, L. Marleau,
Phys. Rev. D{\bf 82}, 054023 (2010).
\bibitem{marleau2}
E. Bonenfant, L. Harbour, L. Marleau,
Phys. Rev. D{\bf 85}, 114045 (2012).
\bibitem{marleau3} 
M.-O. Beaudoin, L. Marleau, arXiv:1305.4944. 
\bibitem{BPS-Sk-nuc1}
C. Adam, C. Naya, J. Sanchez-Guillen, A. Wereszczynski,
Phys. Rev. C{\bf 88},  054313 (2013).
\bibitem{BPS-Sk-nuc2}
C. Adam, C. Naya, J. Sanchez-Guillen, A. Wereszczynski,
Phys. Rev. Lett. {\bf 111}, 232501 (2013).
\bibitem{BPS-Sk-rop}
C. Adam, C. Naya, J. Sanchez-Guillen, A. Wereszczynski,
Phys. Lett. B {\bf 726}, 892 (2013).
\bibitem{harland}
D. Harland, arXiv:1311.2403.
\bibitem{manton}
N. S. Manton, 
Commun. Math. Phys. {\bf 111}, 469 (1987).
\bibitem{self-dual}
C. Adam, L. A. Ferreira, E. da Hora, A. Wereszczynski, W. J. Zakrzewski,
JHEP {\bf 1308}, 062 (2013). 
\bb{krusch}
D. Foster, S. Krusch, J. Phys. A {\bf 46}, 265401 (2013).
\bb{deser}
S. Deser, M. J. Duff, C. J. Isham, Nucl. Phys B{\bf 114}, 29 (1976).
\bb{nicole}
D. A. Nicole, J. Phys. G{\bf 4}, 1363 (1978). 
\bb{nico1}
C. Adam, J. Sanchez-Guillen, R.A. Vazquez, A. Wereszczynski,
J. Math. Phys. {\bf 47}, 052302 (2006).
\bibitem{nico2}
M. Gillard, P. Sutcliffe,
 J. Math. Phys. {\bf 51}, 122305 (2010).
\bb{AFZ1}
H. Aratyn, L.A. Ferreira, A.H. Zimerman,
Phys. Lett. B{\bf 456}, 162 (1999).
\bibitem{AFZ2} 
H. Aratyn, L.A. Ferreira, A.H. Zimerman,
Phys. Rev. Lett. {\bf 83}, 1723 (1999).
\bb{speight3}
J. M. Speight, J. Geom. Phys. {\bf 60}, 599 (2010).
\bb{canfora}
F. Canfora, Phys. Rev. D{\bf 88}, 065028 (2013).
\bibitem{old1} B.M.A.G. Piette, B.J. Schoers and W.J.
Zakrzewski, Z. Phys. C {\bf 65}, 165 (1995).
\bibitem{old2}
 B.M.A.G. Piette, B.J. 
Schoers and W.J. Zakrzewski, Nucl. Phys. B {\bf 439}, 205 (1995).
\bibitem{new} T. Weidig, Nonlinearity {\bf 12}, 1489 (1999).
\bibitem{foster}
D. Foster, Nonlinearity {\bf 23}, 465 (2010).
\bibitem{qhe1} S.L. Soundhi, A. Karlhede, S.A. Kivelson, E.H.
Rezayi, Phys. Rev. B {\bf 47} (1993) 16419.
\bibitem{qhe2}
 N.R. Walet, T. Weidig,
arXiv: cond-mat/0106157.
\bibitem{brane1}
Y. Kodama, K. Kokubu, N. Sawado, Phys. Rev. D{\bf 79}, 065024 (2009).
\bibitem{brane2}
Y. Brihaye, T. Delsate, N. Sawado, Y. Kodama, Phys. Rev. D{\bf 82}, 106002 (2010).
\bibitem{brane3}
T. Delsate, N. Sawado,  Phys. Rev. D{\bf 85}, 065025 (2012).
\bibitem{akama}
K. Akama, "Pregeometry" in: Lecture Notes in Physics, 176, Gauge
Theory and Gravitation, Proceedings, Nara, 1982, ed. K. Kikkawa,
N. Nakanishi and H. Nariai, 267-271 (Springer-Verlag,1983),
hep-th/0001113.
\bibitem{rubakov}
V.A. Rubakov and M.E. Shaposhnikov, Phys. Lett. B{\bf 125}, 136 (1983).
\bibitem{sundrum}
L. Randall, R. Sundrum, Phys. Rev. Lett. {\bf 83}, 4690 (1999).
\bibitem{csaki}
C. Csaki, J. Erlich, T.J. Hollowood, Y. Shirman,
Nucl. Phys. B{\bf 581}, 309 (2000).
\bibitem{poly}
A. A. Belavin, A. M. Polyakov, JETP Lett. {\bf 22}, 245 (1975). 
\bibitem{ward}
M. de Innocentis, R. S. Ward, Nonlinearity {\bf 14} (2001) 663.
\bb{restr-bS}
C. Adam, T. Romanczukiewicz,  J. Sanchez-Guillen, A. Wereszczynski, 
Phys. Rev. D{\bf 81}, 085007 (2010).
\bb{Sp1}
J. M. Speight, 
J. Phys. A{\bf 43}, 405201 (2010). 
\bibitem{sutcliffe1}
P. Sutcliffe, JHEP {\bf 1008}, 019 (2010).
\bibitem{sutcliffe2}
P. Sutcliffe, JHEP {\bf 1104}, 045 (2011).
\bibitem{ma1}
Y-L. Ma, Y. Oh, G-S. Yang, M. Harada, H. K. Lee, B-Y. Park, M. Rho,
Phys. Rev. D{\bf 86}, 074025 (2012). 
\bibitem{ma2}
Y-L. Ma, G-S. Yang, Y. Oh, M. Harada, 
Phys. Rev. D{\bf 87}, 034023 (2013).
\bb{cristal1}
L. Castillejo, P. S. J. Jones, A. D. Jackson, J. J.
M. Verbaarschot and A. Jackson, 
Nucl. Phys. A {\bf 501} (1989) 801.
\bibitem{cristal2}
 M. Kugler and S. Shtrikman, 
Phys. Lett. B 208 (1988) 491.
\bb{speight2}
J. M. Speight, 
arXiv:1307.3063.


\end{thebibliography}
\end{document}